\documentstyle[twocolumn,aps,graphicx]{revtex}

\input epsf

\begin{document}
\title{Linking Ultracold Polar Molecules}
\author{A. V. Avdeenkov and John L. Bohn\cite{byline}}
\address{JILA and Department of Physics,
University of Colorado, Boulder, CO 80309-0440}
\date{\today}
\maketitle

\begin{abstract}
We predict that pairs of polar molecules can be weakly bound together
in an ultracold environment, provided that a dc electric field is
present.  The field that links the molecules together also strongly
influences the basic properties of the resulting dimer, such as its 
binding energy and predissociation lifetime.  Because of their long-range character
these dimers will be useful in disentangling cold collision dynamics
of polar molecules.  As an example, we estimate the microwave
photoassociation yield for OH-OH cold collisions.
\end{abstract}


\narrowtext

A recurring theme that has emerged in the study of ultracold matter
has been one of control over interparticle interactions.  Most
notably, magnetic fields \cite{Ketterle,Courteille,Roberts,Strecker} and 
resonant laser light \cite{Lett} have been employed
to alter atomic scattering lengths, influencing in turn the behavior
of Bose-Einstein condensates.  Perhaps even more intriguing is the
ability, using the same tools, to catalyze the production of
diatomic molecules \cite{Stwalley,Knize,Pillet,Heinzen,McKenzie,Wieman}.  
When the initial atomic sample is Bose condensed, there can result coherent 
superpositions of atoms and molecules \cite{Wieman} in a process dubbed 
``superchemistry'' \cite{Heinzen2}.  More broadly, external fields
may be expected to guide chemical reactions.  Chemical processes
involving atom exchange have been considered at ultralow temperatures
\cite{Esry,Bala,Vardi,Hutson}, and have been observed so far
experimentally in the context of three-body recombination \cite{Ketterle,Wieman}
and vibrational quenching \cite{Heinzen}.

A prerequisite to effective control over atoms or molecules is a clear
understanding of their interactions at ultralow temperatures.  Even for
alkali atoms, {\it ab initio} potentials have insufficient
accuracy to predict scattering lengths; this statement is even more
true for molecules.  Investigators must therefore rely on key experimental
data that will unravel threshold scattering properties of molecules
in the most straightforward way. For the alkali atoms, an efficient 
tool for extracting scattering lengths has been optical photoassociation 
(PA) to weakly-bound states of alkali dimers.
The states in question are generated by spin-orbit couplings that
lead to avoided crossings in potential curves at large values $R$ of interatomic
separation (typically tens of Bohr radii) \cite{Stwalley}.  
Because of their long-range character, they are insensitive
to details of the small-$R$ potentials that are poorly characterized.
Free-to-bound optical spectra of these states therefore contain, in a
reasonably accessible form, important information on scattering
phase shifts \cite{Weiner}.

More recently, even more exotic molecular states have been predicted to
exist in ultracold environments.  Greene and collaborators have predicted
Rydberg states of diatomic molecules with elaborate electronic wave
functions, whose density plots resemble trilobites \cite{Greene} or 
butterflies \cite{Hamilton}.  In addition, C\^{o}t\'{e}, 
Kharchenko, and Lukin have identified possible molecules composed
of hundreds of neutral atoms clinging to a single impurity ion
in a Bose-Einstein condensate \cite{Cote}.

In this Letter we introduce another set of unusual molecular states.
These are composed of two ground-state polar molecules held at large
intermolecular separation under the joint influence of electric dipole
forces and an external electric field.  The properties of these metastable
dimers, such as binding energy, intermolecular separation, and predissociation 
lifetime, depend strongly on the value of the field, leading to
new types of control over intermolecular interactions.
These states may be populated by rapid electric field ramps,
or else by the analog of photoassociation spectroscopy in the 
microwave regime.  Below we describe the origin and properties of
these states,  and estimate the
trap loss rates associated with their formation and subsequent dissociation in
microwave PA experiments.  In this way we sketch a general strategy 
for understanding and manipulating cold collisions of
polar molecules produced by either Stark deceleration \cite{Meijer} or buffer-gas
cooling \cite{Doyle}.

For concreteness we will consider dimers of two OH molecules, 
to be denoted [OH]$_2^*$, although the phenomenon should apply quite 
generally to polar molecules, for example NH$_3$ \cite{Meijer},
that suffer a linear Stark effect at experimentally realizable fields.
We have previously described  a model for the OH-OH interaction Hamiltonian, including
the parity, spin, and nuclear spin degrees of freedom that are most relevant
at ultralow temperatures \cite{Avdeenkov}.  The OH ground state has
spin $j=3/2$, which, coupled to the nuclear spin $i=1/2$ of the
proton, gives total spin $f=2$ in the molecular states we consider.
For clarity, we restrict the present model to molecules that are spin 
polarized with $m_f=f=2$, and neglect couplings to different values of $m_f$.  
Further, we incorporate only the partial
waves $L=0,2$.  These restrictions do not change the qualitative
behavior of the model \cite{Avdeenkov}, but they do render its structure
more tractable. 

The scattering thresholds in this model are set by the Stark splitting
$\Lambda ({\cal E})$ of the molecular levels.  At the relatively low fields 
of interest here, rotational states are not substantially mixed by the field, but
the opposite parity states of the molecule's Lambda doublet are mixed.
Thus there is a weak tendency for the field to align the dipole
moments, along the field when $\omega <0$, and against the field
when $\omega>0$, where $\omega$ is the projection of $j$ along the
body axis of the molecule (these alignments are reversed when the
laboratory projection $m_j<0$) \cite{Schreel}.  For convenience
we refer to these alignments, valid at infinitely large intermolecular separation, 
as $\uparrow$ and $\downarrow$, respectively.
A set of adiabatic potential energy curves for our model is shown
in Figure 1(a), where the dc electric field strength is taken to be
${\cal E}=10^4$ V/cm.

\epsfxsize = 3in
\begin{figure}
\epsfbox{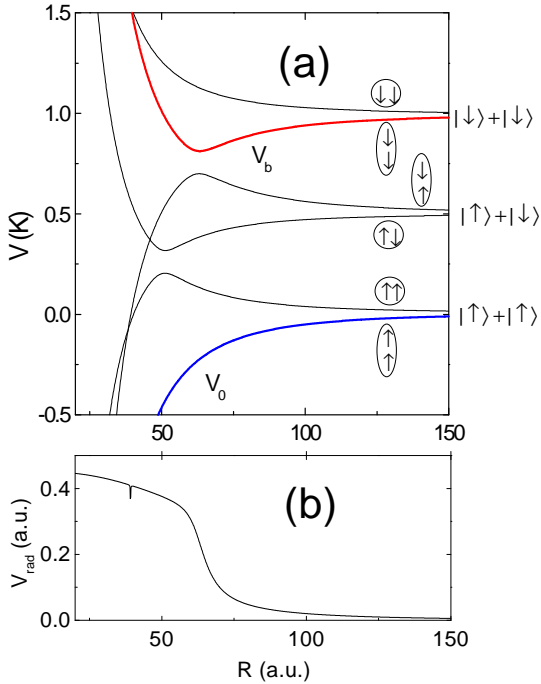}
\vskip .5in
\caption{(a) Adiabatic potential energy curves for the six-channel model
of [OH]$_2^*$, as described in the text.  Avoided crossings in these curves create
the potential well labeled $V_b$, which holds bound states of the dimer. (b)
shows the electric-dipole coupling matrix element between $V_b$ and the 
ground state $V_0$.}
\end{figure}

When the separation between the molecules is not infinite, the partial wave $L$
is no longer a good quantum number, different values being mixed by strong 
anisotropies of the dipole-dipole interaction.  The adiabatic channels shown 
therefore represent different orientations of the intermolecular axis
relative to the laboratory-fixed electric field axis,
as indicated schematically by the diagrams within 
ellipses.  For example, consider the highest threshold where both molecules 
point against the field.  The attractive adiabatic potential corresponds
to a combination of $L=0$ and $2$ partial waves that places the dipoles
end-to-end on average, whereas the repulsive curve places the
dipoles side-by-side on average.  

When the molecules approach still closer to one another, there is an
inevitable curve crossing between the repulsive potential correlating to
the $|\uparrow \rangle + $ $|\downarrow \rangle$ threshold and the 
attractive potential correlating to the $|\downarrow \rangle +$
$|\downarrow \rangle$ threshold.  These channels are dipole-coupled,
leading to an avoided curve crossing at $R=60$ a.u. in the figure.
The resulting potential curve, labeled $V_b$ in Figure 1(a), is deep enough 
to hold a set of quasi-bound states of the [OH]$_2^*$ dimer. (Note
that this curve was denoted ``$U_0$'' in Ref. \cite{Avdeenkov}. Here
we label it more consistently with the theory of photoassociation, to which
we refer below.)

The potential $V_b$, and its bound states, are strongly subject to
the strength of the applied electric field ${\cal E}$.  First, changes
in ${\cal E}$ shift the relative thresholds, changing the location of
the crossing point.  Second, the coupling matrix element between the
channels that cross to form $V_b$ varies with field as $\sim k/(1+k^2)$, where 
$k = 2 \mu {\cal E}/ \Lambda (0)$, and $\mu$ and 
$\Lambda (0)$ are the molecule's dipole moment and zero-field Lambda-doublet 
splitting, respectively \cite{Avdeenkov}.  This means that in low field the 
crossing is purely diabatic, to lowest order, and the molecules
can approach to small values of $R$ where hydrogen bonding
and chemical forces can act.   

That the field controls these states is illustrated in Figure 2.
Here the binding energies of potential $V_b$ are plotted as a function 
of electric field ${\cal E}$, from zero to $10^4$ V/cm.  The solid lines 
refer to a close-coupling calculation in the six-channel model,
whereas the dashed lines give the bound state energies of the
adiabatic potential $V_b$.  Similar results are obtained 
in the more complete calculation described in Ref. \cite{Avdeenkov}.
A striking feature of this figure is that for fields below
$\sim 1000$ V/cm there are no bound states at all, because the
potential becomes too shallow.  Because of the essential role the field
plays in creating these states, we refer to them as ``field linked'' states.

The field-linked (FL) states remain coupled to channels with lower-energy 
thresholds, and can therefore predissociate to these channels.  For 
several selected field values the energy widths for predissociation,
computed from the eigenphase sums in the close-coupling calculation,
are indicated by error bars on the energies.  Two general trends are 
readily apparent: first, the widths as a whole become exponentially
narrower as the binding energy approaches threshold; second, there is an
additional field-dependent modulation in the widths, governed by the
varying overlap integrals between the resonant states and the
continua into which they decay.

The dominance of the electric field over the FL states implies
that ultracold molecular collisions can be manipulated.  For example,
resonant control over scattering lengths in $| \downarrow \rangle + $
$| \downarrow \rangle$ collisions is possible by tuning bound states
of $V_b$ through the scattering threshold \cite{Avdeenkov}.  A similar
tuning was noted for alkali atoms in extremely strong electric fields
\cite{You}, but without making the explicit connection to bound states.

The FL states also imply a degree of control over
chemical reactions.  For example, placing two OH molecules in a FL
state holds them at a ``safe'' distance where the reaction cannot
occur (assuming for example that the exoergic reaction 
OH + OH $\rightarrow$ H$_2$O + O  actually occurs  at
ultralow temperatures).  The reaction can then be activated at will
by switching off the field, which could be done at any desired rate
simply by varying the field strength.

\epsfxsize = 3in
\begin{figure}
\epsfbox{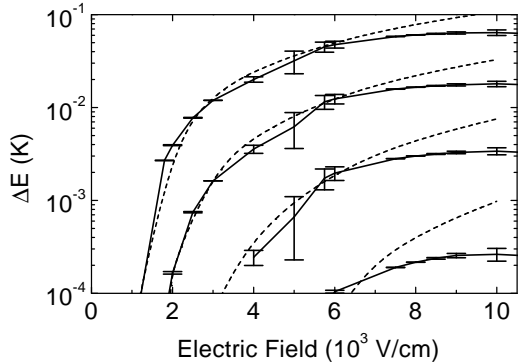}
\vskip .5in
\caption{Binding energies of bound states in the potential $V_b$ versus 
electric field. Solid lines: close-coupling results. dashed lines: 
single-channel adiabatic appoximation. Error bars: approximate predissociation
widths, as computed form the eigenphase sums in the close-coupling
calculations.}
\end{figure}

FL states can presumably be created from a gas of $|\downarrow \rangle$
molecules by quickly ramping the electric field, just as bound states
of Rb$_2$ have been created in magnetic field ramps \cite{Wieman}.
Setting the field to an appropriate value will guarantee that the state
created will be long-lived against predissociation.  It is also 
possible that they have been produced already by collisions in the
supersonic expansion that feeds the Stark slowing process \cite{Meijer},
although they have not yet been sought in this environment.

As a practical issue, a gas of molecules in the weak-electric-field-seeking
$| \downarrow \rangle$ state is subject to large collisional losses in an
electric field
and is therefore not suitable for evaporative cooling \cite{Avdeenkov}.
The $|\uparrow \rangle$ state, on the other hand, is collisionally stable,
being the lowest energy state of the OH molecule, and it can be
trapped in either a magnetostatic or an off-resonant optical dipole trap.
Adiabatic curves correlating to the $|\uparrow \rangle +$ $|\uparrow \rangle$
threshold do not experience a curve crossing at large $R$, and allow the molecules
to approach to small values of $R$ (see Fig. 1).  The cold collision
parameters, for instance the scattering length, will then have to be
measured in a suitable experiment.

We are therefore led to consider photoassociation (PA) spectroscopy
of the molecular pair.  This will be PA at microwave, rather than
optical, wavelengths, appropriate to the energies of the FL
states.  We follow here the discussion of Ref. \cite{Bohn},
and consider single-channel processes only.  This will enable 
order-of magnitude estimates of the multichannel response.  The PA
process may be viewed as a scattering event, where two molecules
incident in channel $V_0$ with relative energy $E$ are promoted by a microwave 
photon of detuning $\Delta$ to a FL state in potential $V_b$.  This 
state can then predissociate, typically releasing a large fraction of the
photon's energy and leading to trap loss in the exit channel.  The rate 
constant for collisional trap loss via a bound state in potential $V_b$
is given by \cite{Bohn}
\begin{equation}
\label{rate}
K_{\rm loss} = {\pi v \over k^2} {\gamma \Gamma \over
(E-\Delta)^2 + ((\gamma + \Gamma) / 2)^2 },
\end{equation}
where $v$ and $k$ are the relative incident velocity and wave number,
$\gamma$ is the predissociation width, and the stimulated
width $\Gamma$ is given by the Fermi's Golden Rule expression
\begin{equation}
\label{width}
\Gamma = 2 \pi |\langle f_0 | V_{\rm rad} | \phi_b \rangle |^2.
\end{equation}
Here $V_{\rm rad}=<-{\vec \mu}\cdot {\vec {\cal E}_{ac}}>$ is the 
$R$-dependent dipole matrix element for the transition, averaged over 
all coordinates except $R$ \cite{Mies}.  ${\cal E}_{ac}$ is the amplitude
of the ac electric field of the applied microwaves, which we assume
are polarized parallel to the dc field.  The width
(\ref{rate}) also depends on $f_0(R)$, the energy-normalized incident wave 
function, and $\phi_b(R)$, the unit-normalized bound state wave function 
in potential $V_b$.  On resonance, and in the typical case $\Gamma \ll 
\gamma$ the rate is approximately
\begin{equation} 
\label{approx}
K_{\rm loss} \sim {4 \pi v \over k^2} {\Gamma \over \gamma},
\end{equation}
hence is proportional to the microwave intensity.

Measurements of optical PA spectra have proven useful in  assessing the
scattering lengths of alkali atoms \cite{Cote2}, largely because
the overlap integral in (\ref{width}) is sensitive to the amplitude
of the wave function $f_0$ at the intermolecular distance $R_c$ where the 
transition occurs.  In fact, this circumstance enables PA rates to be estimated
using a ``reflection approximation'' that focuses on the wave functions
at $R=R_c$ alone \cite{Julienne}.  The FL states should exhibit a
similar sensitivity to scattering length,
although now the $R$-dependence of the dipole coupling $V_{\rm rad}$ cannot
be neglected, as seen in Figure 1(b).
In the limit of large $R$, $V_{\rm rad}$ tends to zero, owing to the
inability of a single photon to drive a transition of both molecules
at once.   At smaller values of $R$, where the avoided crossing occurs 
that creates $V_b$, the radiative coupling attains a nonzero value.

To illustrate the feasibility of PA experiments in this system, 
and to emphasize the strong sensitivity of the spectra to electric
field and scattering length, we estimate rate constants for 
several different circumstances.  These results are summarized 
in Table I.  In all cases we have assumed that the ac microwave field
has a field strength of ${\cal E}_{ac} =15$ V/cm, which can be
reasonably achieved in the laboratory \cite{Ye}.  We have computed
the resonant PA rate according to Eqn. (\ref{rate}), assuming two
different short-range potentials that produce the scattering lengths
$a=69$ a.u. and $a=-400$ a.u. in the incident $V_0$ ground channel,
in zero electric field.  The collision energy is 100 $\mu$K in the
results shown here.

The Tables show, for several values of the dc electric field ${\cal E}$,
the binding energy of several states; their predissociation widths $\gamma$;
their {\it stimulated} widths $\Gamma$ for each of the 
scattering lengths; and the resonant loss rate $K_{\rm loss}$, again
for both scattering lengths.   The table illustrates that
a fairly small field change, on the order of 1 kV/cm, can completely alter the
loss rates.  Moreover, the rates are different for different scattering lengths,
implying sensitivity to details of the short-range part of $V_0$. A complete
map of loss versus field ought therefore to contain a wealth of information
about cold collisions of OH molecules.  Equally importantly, the absolute 
magnitudes of the loss rates can be quite large, as high as several
$10^{-11}$ cm$^3$/sec, which should be easily measurable if 
densities are high enough for molecular collisions to be observed at all.

In summary, we have introduced a set of weakly-bound ``field-linked''
states of ultracold polar molecules that exist only in the presence 
of an electric field.  The sensitivity of these states to electric
field suggests an unprecedented level of control over intermolecular
forces, since the very shape of the interaction potential can be
manipulated.  In addition, their purely long-range character
makes them invaluable tools for approaching the difficult problem
of deciphering ultracold molecular collisions.

This work was supported by the NSF.  We acknowledge useful discussions
with J. Hutson and J. Ye.

\begin{table} 
\caption{Characteristics of field-linked states and the photoassociation yield 
subject to an ${\cal E}=15 V/cm$ ac electric field, at a collision energy
$E = 100$ $\mu$K.  Numbers in parentheses denote the power of 10 by which the
numbers are to be multipled.} 
\begin{tabular}{|c|c|c|c|c|c|c|}  
 ${ \cal E}$ &  $\Delta E$ &   \multicolumn{2}{c|}{$\Gamma$}  & $\gamma$ &  \multicolumn{2}{c|}{$K_{\rm loss}$} \\  
  { kV/cm} &  {\scriptsize K} &   \multicolumn{2}{c|}{ $\mu$ K}  & { $\mu$ K } & \multicolumn{2}{c|}{$cm^{3}/sec$} \\ [0.02cm] \cline{1-7}   
 {0}&{}&{a=69}&{a=-400}&{}&{a=69}&{a=-400}\\  \cline{1-7} 
   {2}    &   {4.3(-3)}   &   {4.3(-1)}    &   {1.1(-1)}    &   {90.9}   &   {7.4(-12)}   &   {1.9(-12)} \\ [0.02cm] 
   {}     &   {2.1(-4)}   &   {3.6(-1)}    &   {6.5(-3)}    &   {10.1}   &   {5.3(-11)}   &   {1.0(-12)} \\ [0.02cm] \cline{1-7} 
   {4}    &   {2.5(-2)}   &   {1.3(-2)}    &   {1.6(-1)}    &   {2691.2} &   {7.4(-15)}   &   {9.3(-14)} \\ [0.02cm] 
   {}     &   {5.1(-3)}   &   {1.5(-3)}    &   {1.5(-1)}    &   {719.3}  &   {3.2(-15)}   &   {3.3(-13)} \\ [0.02cm] 
   {}     &   {4.5(-4)}   &   {3.3(-7)}    &   {5.4(-2)}    &   {90.0}   &   {5.8(-18)}   &   {9.5(-13)} \\ [0.02cm] \cline{1-7} 
   {8}    &   {2.3(-2)}   &   {12.0}       &   {4.4(-2)}    &   {225.7}  &   {7.6(-11)}   &   {3.1(-13)} \\ [0.02cm] 
   {}     &   {4.6(-3)}   &   {5.2(-3)}    &   {1.3(-2)}    &   {29.6}   &   {2.8(-13)}   &   {6.9(-13)} \\ [0.02cm] 
   {}     &   {4.4(-4)}   &   {1.3(-3)}    &   {3.5(-3)}    &   {10.6}   &   {1.9(-13)}   &   {5.2(-13)} \\  [0.02cm] \cline{1-7} 
   {10}   &   {3.3(-2)}   &   {7.2(-5)}    &   {1.1(-2)}    &   {8850.1} &   {1.3(-17)}   &   {1.9(-15)} \\ [0.02cm] 
   {}     &   {7.5(-3)}   &   {1.6(-4)}    &   {4.8(-3)}    &   {2383.0} &   {1.0(-16)}   &   {3.2(-15)} \\ [0.02cm] 
   {}     &   {9.8(-4)}   &   {6.8(-5)}    &   {1.4(-3)}    &   {580.2}  &   {1.9(-16)}   &   {3.9(-15)} \\  [0.02cm] \cline{1-7}    
\end{tabular}  
\end{table}

\end{document}